\RequirePackage{lineno}
\documentclass[reprint]{revtex4-2} 
\usepackage{graphicx}
\usepackage{dcolumn}
\usepackage{bm}
\usepackage{color}
\usepackage{caption}
\usepackage{amsmath}
\usepackage{caption} 
\usepackage{amssymb} 

\usepackage[T1]{fontenc} 
\usepackage[utf8x]{inputenc}

\makeatletter 
\newcommand*{\balancecolsandclearpage}{%
  \close@column@grid
  \cleardoublepage
  \twocolumngrid
}

\makeatother

\begin{document}
\preprint{}

\title{Single Crystalline 2D Material Nanoribbon Networks for Nanoelectronics}

\author{Muhammad Awais Aslam{$^{1}$}}\email{muhammad.aslam@unileoben.ac.at}
\author{Tuan Hoang Tran$^{2}$}
\author{Antonio Supina$^{3}$}
\author{Olivier Siri$^{4}$}
\author{Vincent Meunier$^{5}$}
\author{Kenji Watanabe$^{6}$}
\author{Takashi Taniguchi $^{7}$}
\author{Marko Kralj$^{3}$}
\author{Christian Teichert$^{1}$}
\author{Evgeniya Sheremet$^{2}$}
\author{Raul D. Rodriguez$^{2}$}
\author{Aleksandar Matkovi\'{c}$^{1}$}\email{aleksandar.matkovic@unileoben.ac.at}

\affiliation{$^1$ Institute of Physics, Montanuniversit\"at Leoben, Franz Josef Strasse 18, 8700 Leoben, Austria}
\affiliation{$^2$ Tomsk Polytechnic University, Lenina ave. 30, 634034, Tomsk, Russia}
\affiliation{$^3$ Surfaces, Interfaces and 2D Materials Research Group, Institute of Physics, Bijenička cesta 46, 10000 Zagreb, Croatia}
\affiliation{$^4$ Aix Marseille University CNRS CINaM UMR 7325 Campus de Luminy 13288, Marseille cedex 09, France}
\affiliation{$^5$ Department of Physics, Applied Physics, and Astronomy, Rensselaer Polytechnic Institute, Troy, New
York 12180, United States}
\affiliation{$^6$ Research Center for Functional Materials, National Institute for Materials Science, 1- 1 Namiki, Tsukuba 305-0044, Japan}
\affiliation{$^7$ {International Center for Materials Nanoarchitectonics, National Institute for Materials
Science, 1-1 Namiki, Tsukuba 305-0044, Japan}}

\date{\today}

\begin{abstract}
\textbf{The last decade has seen a flurry of studies related to graphene nanoribbons owing to their potential applications in the quantum realm. However, little experimental work has been reported towards nanoribbons of other 2D materials due to the absence of synthesis routes. Here, we propose a universal approach to synthesize high-quality networks of nanoribbons from arbitrary 2D materials while maintaining high crystallinity, sufficient yield, narrow size distribution, and straight-forward device integrability. The wide applicability of this technique is demonstrated by fabricating MoS$_2$, WS$_2$, WSe$_2$, and graphene nanoribbon field effect transistors that inherently do not suffer from interconnection resistances. By relying on self-assembled and self-aligned organic nanostructures as masks, we demonstrate the possibility of controlling the predominant crystallographic direction of the nanoribbon's edges. Electrical characterization shows record mobilities and very high ON currents for various TMDCs despite extreme width scaling. Lastly, we explore decoration of nanoribbon edges with plasmonic particles paving the way towards the development of nanoribbon-based plasmonic sensing and opto-electronic devices.}

\end{abstract}

\maketitle 


\textbf{Introduction}

The successful synthesis of graphene nanoribbons (NRs)\cite{li2008chemically,cai2014graphene} and their implementation in devices\cite{chen2020graphene, saraswat2021materials} has brought them at the forefront as building blocks for information processing in quantum and classical electronics.\cite{wang2021graphene} Graphene NRs enable various functionalities including tunable band gap, high current carrying capability, long mean free path, localized spin and topological edge states.\cite{wang2021graphene} Similarly, other 2D material NRs can display unique edge specific properties such as ferromagnetism\cite{pan2012edge,slota2018magnetic}, efficient catalysis\cite{lin2014enhanced,karunadasa2012molecular}, and enhanced sensing abilities.\cite{lihter2021electrochemical,chen2020graphene} Moreover, a recent study about MoS$_2$ NRs demonstrated their potential for spintronics and quantum transport.\cite{li2021nickel} The development of 2D material NRs is largely driven by the needs in nanoelectronics, where three-dimensional (3D) gate-all-around architectures that employ nanotubes, nanorods, or NRs are considered as the likely solution to the arising scaling challenges.\cite{thomas2020gate,chen2020finfet,xu2021can}

Despite all the possibilities that 2D material-based NRs hold, their sufficient quality, narrow widths, density, controlled edges, and high yield remain as technological challenges for realistic applications. The most widely used preparation methods are bottom-up chemical synthesis\cite{jia2011graphene,barone2006electronic,jolly2020emerging,bennett2013bottom} and top-down lithography.\cite{han2007energy,chen2019monolayer,tapaszto2008tailoring} Chemical synthesis of NRs offers excellent edge control; however, the device channels suffer from electrical percolation issues and high junction (node) resistances.\cite{chen2020graphene} Furthermore, bottom-up synthesis routes focus almost exclusively on graphene NRs, facing significant obstacles to develop more complex 2D materials or NR heterostructures. Whereas, top-down lithography-based approaches do not offer a straight forward control over NR's alignment with respect to high symmetry directions of the 2D material. They also cause interface contamination degrading device performance and operation.\cite{zhou2010making}

Recently, Areej et al.\cite{aljarb2020ledge} demonstrated a technique based on vicinal growth to fabricate NRs of arbitrary 2D transition metal dichalchogenides (TMDCs). Although, the NRs produced by this approach are single crystalline, the ribbon widths are rather large and non-uniform. The growth of these ribbons is also dependent on specific substrate, requiring an additional transfer step for their integration.\cite{aljarb2020ledge} Vapour Liquid Solid (VLS) growth is another interesting method utilised for the NR synthesis.\cite{li2018vapour,li2021nickel} Despite the fact that VLS uses SiO$_2$ as a substrate, it employs salt and metal precursors that can be detrimental for device integration.  

\begin{figure*}
\centering
\includegraphics[width=1.0\textwidth]{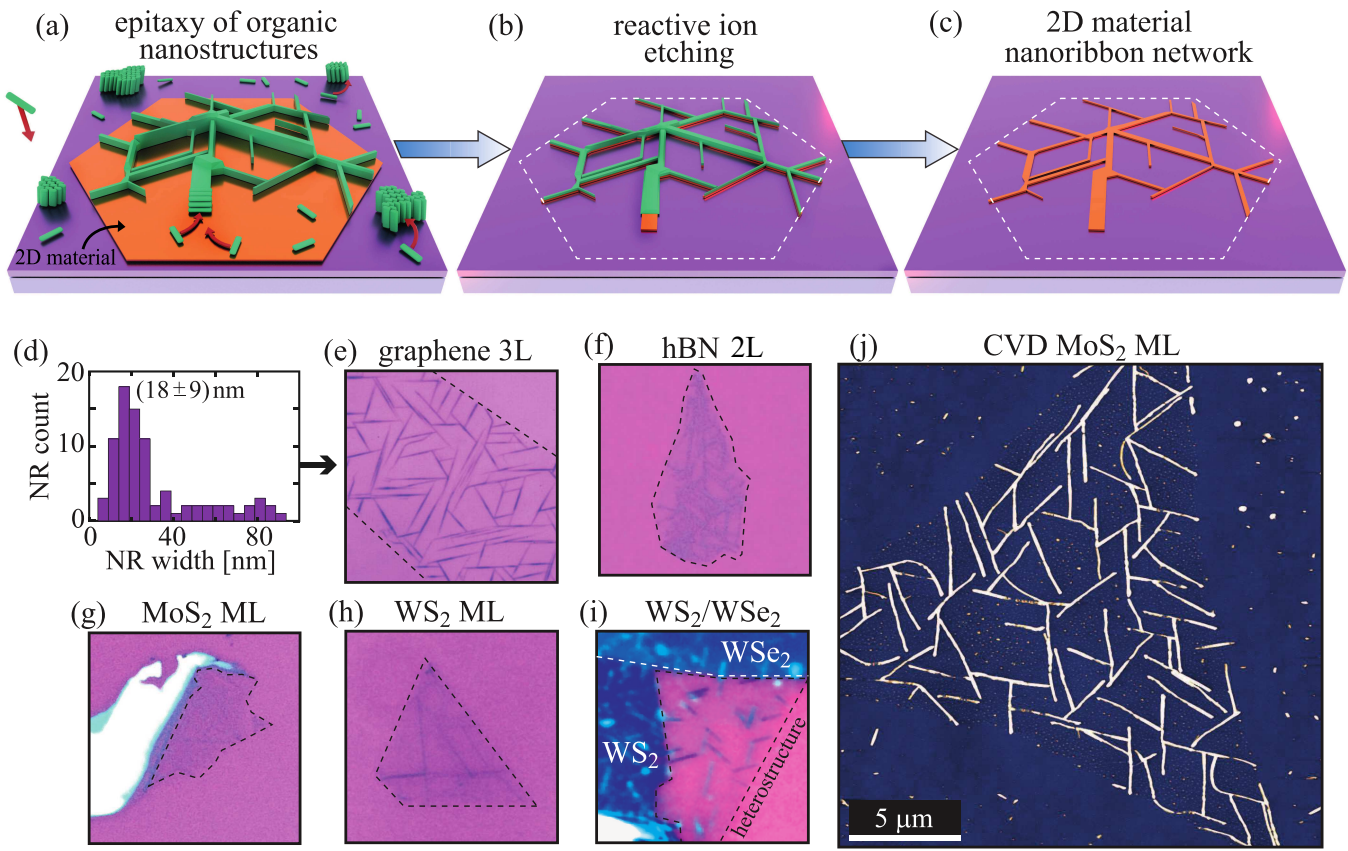}
\caption{Fabrication of 2D material NRNs: (a-c) Schematic representation of the key fabrication steps. (a) Organic nanostructures self-assembly and self-alignment. (b) After reactive ion etching. (c) NRN after removal of the sacrificial organic layer.(d) Histogram of NR widths for a graphene NRN shown in (e). (e-i) Optical micrographs of various 2D material NRNs, presenting respectively NRNs of graphene, hBN, MoS$_2$, WS$_2$, and WS$_2$/WSe$_2$ heterostructure. (j) AFM topography image of a NRN from CVD Monolayer(ML) MoS$_2$ (22$\times$22 $u$m$^2$, $z$-scale 15~nm)}
\label{Figure:1}
\end{figure*}

In this work we tackle the outlined challenges and demonstrate a universal method to fabricate NRs of arbitrary 2D materials, including graphene, hexagonal boron nitride (hBN), transition metal dichalcogenides TMDCs, and nanoribbon heterostructures with a width ranging from 6 to 100~nm. Our approach is based on epitaxially grown organic needle-like nanostructures which self assemble along high-symmetry directions of 2D materials. We exploited these organic nano-needles as a mask through which 2D materials could be etched by oxygen plasma. Resulting are crystalline nanoribbon-networks (NRNs) with high edge-to-surface ratio and controlled predominant crystallographic edge-directions. To investigate the electrical performance of our NRNs and to show a challenging technological application, field effect transistors (FETs) were directly fabricated on Si/SiO$_2$ and Si/SiO$_2$/hBN substrates. Besides their inherent single-crystalline nature, NRN-FETs were obtained without any additional transfer steps. TMDC NRN-based devices show excellent electrical properties, including WS$_2$ and WSe$_2$ nanoribbons which are reported here for the first time. We also observed ferroelectric switching for graphene NR devices due to water adsorption at the ribbon edges.\cite{caridad2018graphene} Our proposed method allows ribbons not to suffer from high node resistances between interconnecting NRs, as the networks are "carved out" from single crystals. To confirm this, we employ \textit{in~operando} Kelvin Probe Force Microscopy (KPFM). To demonstrate our methods scalability and ultimate control over the NR edge-direction, we have fabricated predominant armchair and zigzag NRN from a large-area  MoS$_2$ ML obtained by chemical vapor deposited (CVD). Such high edge-density in NRN-FETs has potential benefits in sensing applications and in tunable catalytic devices, especially when considering catalytic edge reactivity of MoS$_2$.\cite{salazar2020site,wang2019structural} To go one step further, we illustrate edge-specific decoration of NRNs with silver nanoparticles, creating mixed-dimensional plasmonic heterostructures.

\begin{figure}
\centering
\includegraphics[width=0.5\textwidth]{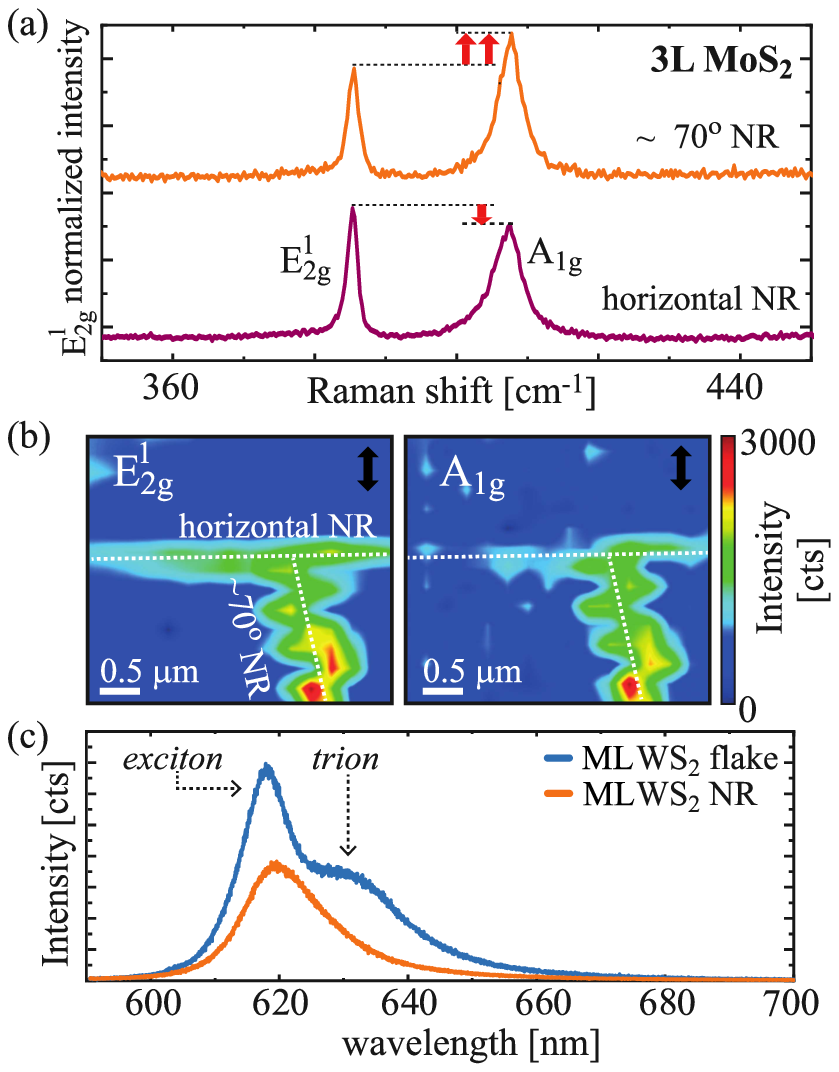}
\caption{Evidence of structural integrity of the NRs. (a) Raman spectra of the characteristic vibrational modes of MoS$_2$ showing the tri-layer flake before patterning, and two ribbons with $\approx$70$^\circ$ of relative inclination. The spectra are normalized with respect to the E$^1$$_{2g}$-mode intensity, and relative changes of the A$_{1g}$ mode are indicated by horizontal dashed lines and red arrows. (b) Raman intensity maps of E$^1$$_{2g}$ and A$_{1g}$ modes of two intersecting NRs, highlighting a distinct variation of A$_{1g}$ mode as the ribbon direction is changed. The direction of linearly polarized light is indicated by black arrows. This is in agreement with the published result that the relative intensity of the A$_{1g}$ to E$^1$$_{2g}$ peaks is strongly dependent on the orientation angle of the material’s crystallographic axes, for a fixed in-plane polarization\cite{liang2014first}.  (c) PL spectra of a ML WS$_2$ flake and corresponding NR, highlighting exciton and trion components. 
}
\label{Figure:2}
\end{figure}

\textbf{2 Results and discussions}

$ $\par
\textbf{Fabrication of NRNs}
$ $\par

Figures~1(a-c) depict the proposed NRN fabrication pathway. Detailed steps for MoS$_2$ and graphene are given in Figure S1 of Supporting Information.  
Typically, NRs of about {10-30~nm} width distribution were realized Figure~1d. By further optimization of the growth and etching time, organic nanostructures mean NRN width can be reduced down to few nanometers.

$ $\par
\begin{figure*}
\centering
\includegraphics[width=1.0\textwidth]{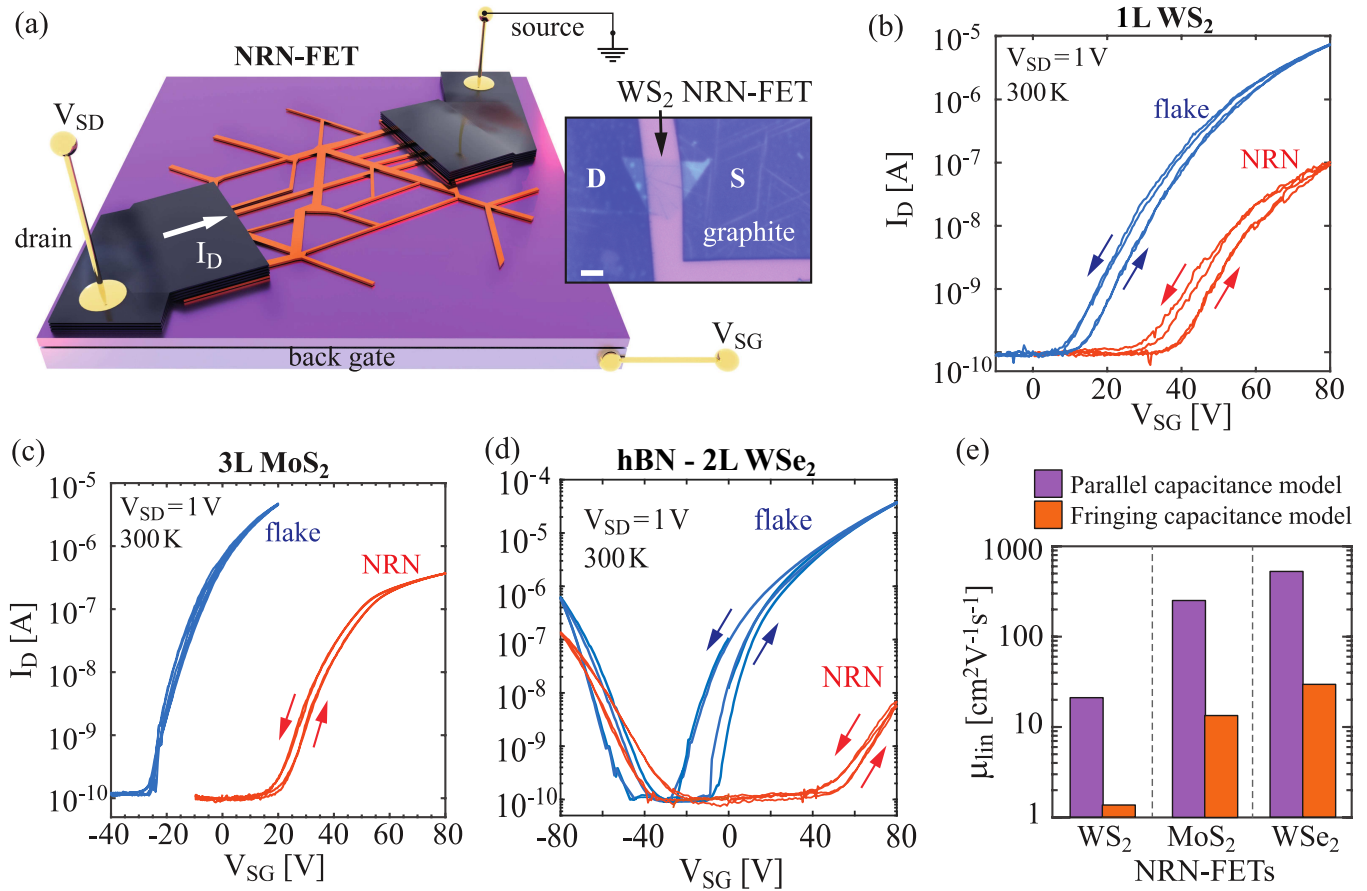}
\caption{TMDC-NRN-FETs (a) Schematic diagram of a NRN-based FET utilizing vdW graphite electrodes. Inset of (a) shows an optical micrograph of a 3L WS$_2$ FET {(scale-bar 5~$\mu$m)}. (b-d) Semi-logarithmic transfer curves of WS$_2$, MoS$_2$, and hBN-WSe$_2$ FETs before and after patterning the flakes into NRNs. (e) Parallel and fringing capacitance apparent linear electron mobilities at 77~K, for the devices presented in (b-d).}
\label{Figure: 3}
\end{figure*}

Organic molecules grow on 2D materials via van der Waals (vdW) epitaxy.\cite{koma1999van} The molecules at the interface with the 2D material substrate adopt a flat-lying orientation and align their $\pi$-networks to optimize vdW interaction with the substrate.\cite{hlawacek2011smooth} Consequently, the molecules at the interface are "locked" into preferential adsorption sites on the substrate and the growing crystallites adopt rotational commensuration to their 2D material support. This provides an inherent self-alignment with the substrates’ high symmetry directions, \textit{i.e.,} armchair or zig-zag.\cite{matkovic2016epitaxy,wang2017probing, vasic2018molecules} By controlling the deposition rate of the molecules and the 2D material substrate temperature, it is possible to tune the length, width, density, and consequently the edge-to-surface ratio of the NRNs.

After organic nanostructure growth, the hybrid organic/2D material stacks were precisely etched to form NRNs via exposure to oxygen plasma, i.e. reactive ion etching (RIE). An etch rate of approximately 1~layer in 3~seconds was established for graphene and TMDCs. Upon etching, the remaining organic molecules can either be left as an encapsulation layer \cite{ha2012hybrid} or removed by rinsing in chloroform or also by vacuum annealing. 
Our method allows for fabrication of monolithic NRNs of different exfoliated and CVD 2D materials, as demonstrated for graphene, hBN, MoS$_2$, WS$_2$ and WSe$_2$ Figures~1(e-j). Apart from individual 2D materials, in Figure~1i we show NRN-heterostructures consisting of vertically stacked monolayer WS2 (n-type) and bilayer WSe2 (p-type), thus enabling atomically thin p-n junctions.

The structural integrity of NRs was probed by Raman and photoluminescence (PL) spectroscopies. 
Results for ML MoS$_2$ and graphene (Figure~S2 in Supporting Information) show no change in the Raman spectra after etching of the flake into NRN. This confirms that the atomic structure integrity of NRs remains intact.  Figures~2(a,b) show results for the NR pairs (considering 3 layer MoS$_2$) with approximately 70$^\circ$ relative inclination. They exhibited a prominent difference in the intensities of E$^1$$_{2g}$ and A$_{1g}$ Raman active modes. Such anisotropy for Raman modes of MoS$_2$ NRs\cite{wu2016monolayer} and MoS$_2$ flakes\cite{ji2016giant} was previously observed by changing the polarization configurations. The intensity of a Raman mode is proportional to the dot product of the Raman tensor with the light polarization. Since the NRs of both directions are etched out of the same flake, their crystallographic orientation, and consequently the dot products are the same. Thus, this anisotropy cannot be explained by the selection rules per se. Raman spectra for rotated nanoribbons and flake are shown in Figure S3 of the Supporting Information. Anisotropy has an apparent effect on the relative A$_{1g}$ mode intensity that warrants further investigation into its variations in narrow NRs. Furthermore, WS$_2$ NRs exhibited a dominant exciton peak and a suppressed trion peak in the PL measurements. This is due to a reduction in the free electron density over small widths (<30nm) caused by the predominant oxygen terminated edges .\cite{cong2018optical,kwon2018variation} An example of the PL spectrum of WS$_2$ is presented in Figure~2(c), comparing the initial ML flake and the resulting NR. The Raman spectra for the nanoribbon WS$_2$/WSe$_2$ heterostructure is provided in the Figure~S4 of the Supporting Information.

\begin{figure*}
\centering
\includegraphics[width=1.0\textwidth]{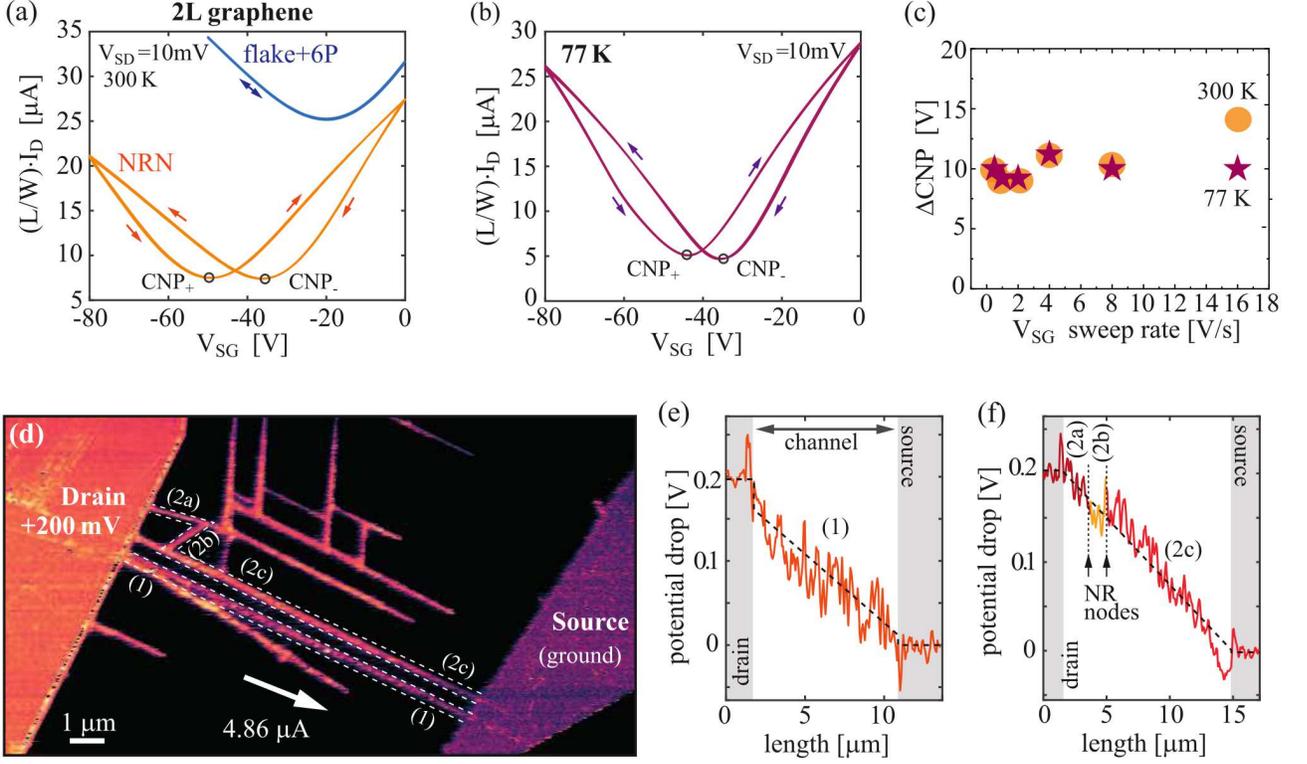}
\caption{Graphene NRN-FETs. (a) Length and width scaled transfer curves of the bi-layer graphene flake after organic nanostructure growth, and of the corresponding NRN-FET after annealing. Forward and reverse sweep direction is indicated by the arrows. (b) Length and width scaled transfer curves for graphene NRN-FET at 77~K, demonstrating that bi-modal switching persists at low temperature. (c) Difference in the forward and reverse bias charge neutrality point (CNP) positions as a function of the V$_{SG}$ sweeping rate. (d) \textit{in operando} FM-KPFM image of a graphene NRN-FET. Potential profile lines from the ribbons indicated by (1) and (2a-c) in sub-panel (d) are presented in (e) and (f), respectively.}
\label{Figure 4}
\end{figure*}

\par$ $\par

\textbf{TMDC-NRN field effect transistors}\par
Field-modulation of the NRNs was tested by fabricating two-terminal field-effect transistors (FETs) and investigating their transfer characteristics (I$_D$(V$_{SG}$)) at 77~K and 300~K. Van der Waals graphite electrodes were employed to probe electrical response of the devices between each step of the fabrication. Our proposed NRN fabrication method is compatible with conventional 2D-FET fabrication schemes (as mask-lithography or e-beam lithography) since the 2D material films can be patterned into NRNs prior to the electrode fabrication. Figure~3a  presents the scheme of the device geometry, and an optical micrograph for one of the WS$_2$-NRN-FETs in the inset.

Figures 3(b-d) provide typical semi-logarithmic transfer curves at 300~K for flakes with organic nanostructures for WS$_2$, MoS$_2$, and WSe$_2$, respectively and after patterning them into NRs. For WSe$_2$ 10 - 15nm hBN was used for bottom capping and NRN was patterned on top of it. MoS$_2$ and WS$_2$ devices exhibit an n–type behaviour whereas WSe$_2$ exhibited an ambipolar behaviour, both before and after patterning of the flakes into NRNs. A shift in the positive V$_{SG}$ direction of the I$_D$(V$_{SG}$) curves was observed after NRN formation, indicating $p$-type doping by the RIE process. On average, I$_D$(V$_{SG}$) curves for MoS$_2$ devices showed a positive shift of 40~V after the NRN formation. This large positive shift can be attributed to electron depletion by the oxygen terminating NR-edges\cite{liu2012hbox}.

The devices exhibited exceptional transfer characteristics even when only SiO$_2$ was used as the gate dielectric. The performance of NRN-FETs could be further enhanced by employing high-$K$ dielectric materials such as HfO$_2$.\cite{lee2013flexible} An increase in the hysteresis was observed in the NR devices (except WSe$_2$) which can be due to two possible mechanisms including high density of edges facilitating charge trapping/de-trapping mechanisms from adosrbates\cite{chen2019monolayer} or increased capacitive gating effect-traps due to SiO$_2$.\cite{park2016thermally} To investigate the origin of hysteresis the devices were annealed in vacuum at 400K which lead to reduction of hysteresis due to removal of adsorbates on the edges\cite{kaushik2017reversible}. However, a full closure of the hysteresis was noted by performing low temperature measurements (77~K) as shown in Figure S5 of the Supporting Information which point towards capacitive gating as the possible reason.\cite{park2016thermally}

Compared to the unetched flakes, the NRN-devices experienced a decrease of the I$_D$, as a consequence of the severely reduced channel widths when compared to the original 2D material-FET.\cite{liu2012hbox} An increase in I$_D$  was observed when NRN-FETs are measured at low temperatures (77~K) as shown in Figure S5 of the Supporting Information. This observation confirms band transport in NRNs. 

Figure~3e summarizes the apparent linear electron mobilities ($\mu$) obtained from the I$_D$(V$_{SG}$) curves measured at 77~K and calculated both by parallel and fringing capacitance models. Mobility plots are shown in Figure S6 of the Supporting Information. The commonly used parallel capacitance model overestimates the mobilities when applied to NRs as their widths are much smaller compared to the oxide thickness.\cite{li2021nickel} Therefore, taking into account the capacitance per unit area for the fringing capacitance model\cite{liao2011thermally}, a more realistic area-specific gate capacitance can be expressed as:

\begin{equation} C_{ox} \approx \varepsilon_{ox} \varepsilon_{0}\left\{\frac{\pi}{\ln \left[6\left(\frac{t_{ox}}{W}+1\right)\right] W}+\frac{1}{t_{ox}}\right\} \end{equation}

where t$_{ox}$ is the oxide thickness and W is channel width. Effective channel length and width were estimated considering parallel and serial connections of NRs for each particular NRN-FET.

\par$ $\par

\textbf{Graphene NRN-FETs and edge-induced ferroelectric effect}\par

\begin{figure*}
\centering
\includegraphics[width=1.0\textwidth]{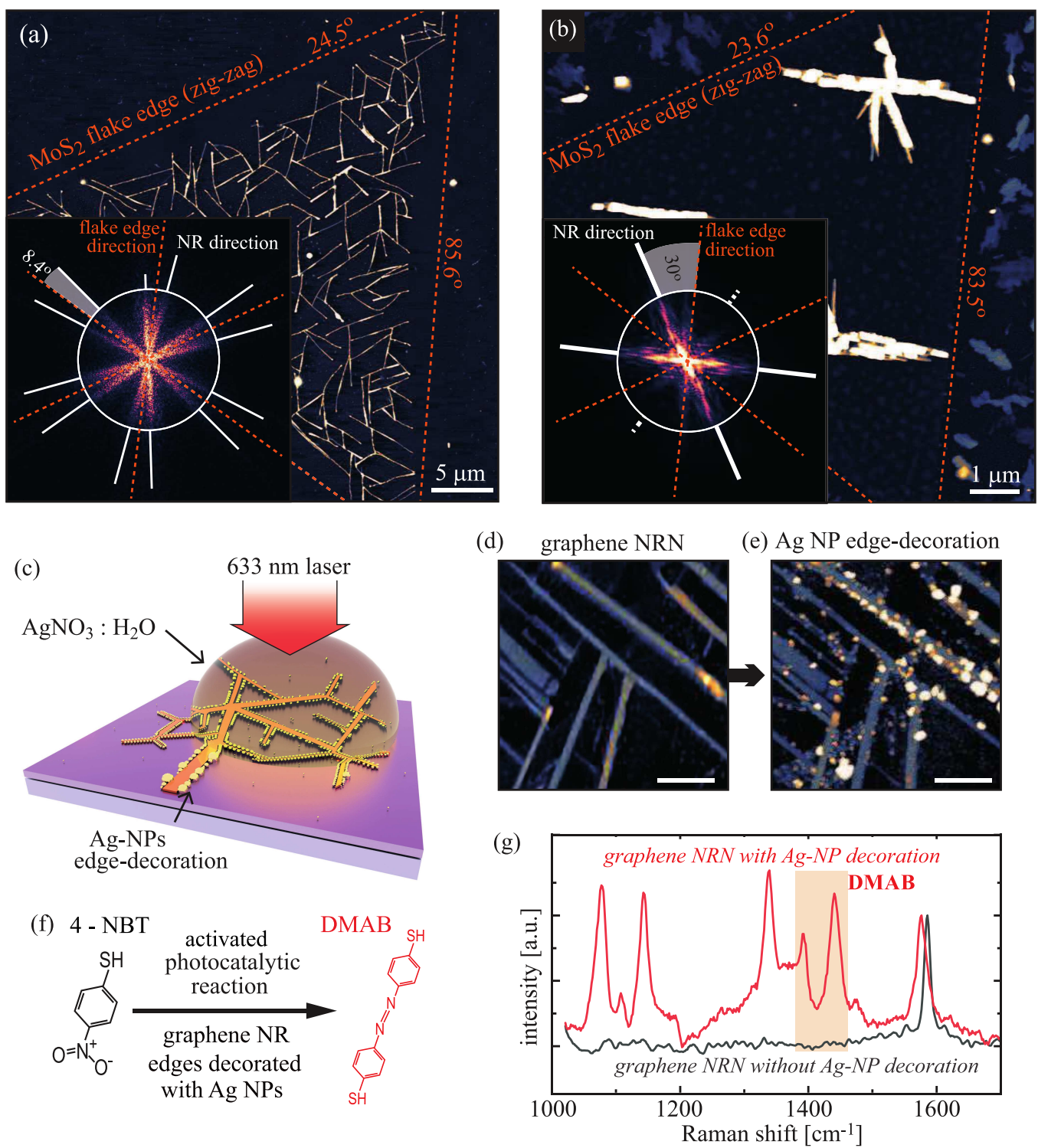}
\caption{\textbf{} (a) CVD MoS$_2$ NRN with ribbons nearly parallel to the flake edges (zigzag direction), using 6P molecules for the self-assembled mask. Inset shows 2D-FFT analysis of the NR directions with respect to the triangular flake edges. FFT image is rotated by 90$^\circ$ to represent real-space directions. (b) Similar to (a) only using DHTA7 molecules for the self-assembled mask, and resulting in NRs in armchair direction (perpendicular to flake edges). (c) Illustration of the process for the decoration of the NRNs with metallic nanoparticles, and (d,e) graphene NRN before and after edge-decoration with Ag NPs (scale bars 500~nm, z-scales 25~nm). (f) Dimerization of 4-NBT to DMAB and (g) demonstration of the photocatalytic activity and SERS capability of the hybrid graphene-NRN + Ag-NP system.}
\label{fig:5}
\end{figure*}

Besides very high mobilities $\approx$ ~1000~-~1200 $\mathrm{cm}^{2} \mathrm{V}^{-1} \mathrm{s}^{-1}$ (using fringing capacitance model) graphene-based NRN devices exhibit pronounced hysteresis in the I$_D$(V$_{SG}$), see Figure 4(a). The observed hysteresis $\Delta$CNP (difference between two CNPs) was not present in the original flakes, nor is introduced by the deposited organic nanostructures. The effect appears after the RIE in oxygen plasma and remains in high-vacuum, at low temperatures Figure 4b, and is practically independent of the V$_{SG}$ sweep-rates Figure 4c. Similar effect was predicted for edge-adsorbed water molecules and was observed for graphene-FETs with oxygen-plasma etched edges.\cite{caridad2018graphene} The orientation of water molecules can be changed due to the torque induced by the external electric field. 
The total field experienced by graphene NRN is a sum of the gate-bias induced field and the net field produced by the edge-adsorbed water dipoles, yielding a robust bi-modal -- ferroelectric -- behavior of graphene NRN-FETs.

To rule out any causes of hysteresis due to trap-states the devices were measured in high vacuum (10$^{-7}$~mbar), after vacuum annealing (at 410~K for over 90~minutes), and were subjected to low temperature (77~K) measurements. In all cases, the observed hysteresis was preserved. As the trapping is sensitive to temperatures a significant quenching would occur at low temperatures.\cite{wang2010hysteresis,singh2017reversible} This was observed for TMDC-based NRN-FETs, where the bi-stable states of the adsorbed water molecules at the ribbon edges are not expected. For graphene NRN-FETs at 77~K the hysteresis of the transfer curves is only slightly reduced as shown in Figure 4(b). In addition, V$_{SG}$ sweep-rate dependent measurements were carried out both at 300~K and at 77~K. Figure~4c presents the V$_{SG}$ sweep-rate dependence of the $\Delta$CNP, representing the negligible difference in the CNP position for the forward and the backward sweeps. This further helps us excluding any contributions from capacitive gating which acts on seconds time scale.\cite{wang2010hysteresis}  Our results point to the induced ferroelectric effect in oxygen-terminated graphene nanoribbon-FETs, which is very similar and more robust than observed previously for the oxygen-terminated flake-edges\cite{caridad2018graphene}. While ferroelectric-graphene nanoribbons and their integration into heterostructures are very interesting and promising pathways for future nanoelectronics, optoelectronics, neuromorphic electronics, and sensing applications, such research is beyond the scope of this study.

To directly probe the resistivity of the nodes between the adjacent NRs and the potential drops across the NRN-FET channel \textit{in operando} frequency modulated (FM) KPFM was performed on graphene NRN-FETs. Figure~4(d) presents a contact potential difference (CPD) map across a graphene NRN-FET during operation. 
To highlight the relevant potential drops across the channel, cross-sections marked in Figure~4d are provided in Figure~4(e,f). The transitions from the electrodes to the channel do not introduce any significant potential drops, as seen earlier for gold electrodes on both graphene and MoS$_2$\cite{yu2009tuning, matkovic2020interfacial}. 
Graphene NR labeled (1) interconnects between the source and drain electrodes, it exhibits an almost perfectly linear potential drop across the 10~$\mu$m long channel Figure~4e. NRs labeled (2a-c) form a parallel connection to NR (1). No potential drop was observed at the nodes between (2a)-(2b) and (2b)-(2c), as indicated in Figure~4f. Further, many more ribbons that do not bridge the source and drain electrodes, maintain a constant potential as these are not part of the current flow across the device. Above all, the consistent potential observed at all nodes is due to the translation of single-crystallinity of the original 2D material into the nanoribbon network.

\par$ $\par
\textbf{Predominant crystallographic orientations of nanoribbons}\par

To demonstrate that our proposed method offers control of NRs predominant crystallographic orientation, two different organic molecules -- parahexaphenyl (6P) and dihydrotetraazaheptacene (DHTA7) -- were grown epitaxially on ML MoS$_2$ obtained by CVD. As their phenylene (6P) and acene (DHTA7) backbones could be seen as armchair and zig-zag motifs, respectively, and upon adsorption, molecular backbones will align with the corresponding high-symmetry directions of the 2D material substrate\cite{kratzer2019adsorption,matkovic2019light,vasic2018molecules}. The control over the orientation of the predominant NR direction can be verified by using triangular CVD MoS$_2$ flakes that terminate with zig-zag edges due to the growth kinetics\cite{van2013grains,zhu2017capture,zhou2013intrinsic}. Figure~5(a,b) compare the NR directions and the triangular MoS$_2$ flake-edge directions, presenting 2D Fast Fourier Transform (2D-FFT) analysis of the AFM topography images (corresponding insets). In the case of 6P masks (Figure~5a), predominant NR directions are tilted by (8.5~$\pm$~0.4)$^\circ$ from the edge directions, \textit{i.e.,} NR edges are close to parallel with the zigzag crystallographic direction. By altering the backbone of the molecular mask (the case of DHTA7 -- Figure~5b) the NR edges change predominantly following the armchair crystallographic direction. Moreover, employing other molecular species could allow controlling this angle for a particular 2D material of interest, and to exploit orientation specific properties for 2D materials\cite{li2019strain, arab2015anisotropic} in the one-dimensional NR-regime.

To demonstrate unique characteristics particularly high edge-to-surface ratio of our 2D material nanoribbon networks, we investigated edge-specific decoration of the NRs by metallic nanoparticles (NPs). Figure~5c schematically presents the decoration process. The details are provided in the methods section. The NR edges induce selective nucleation of Ag nanoparticles via the photo-activated reduction of Ag ions at the edges via electron transfer from graphene. Figure~5d,e present a graphene NRN before and after decoration with Ag NPs. Edge-specific decoration of 2D materials with metallic NPs was already demonstrated\cite{tanaka2009photochemical}, and utilizing NRN enhances the benefits of these hybrid systems.

To investigate the photocatalytic activity of the edge-decorated NPs, we use 4-nitrobenzenethiol (4-NBT)  as a model for photocatalysis experiment. This is shown in Figure~5f. The photocatalytic conversion of 4-NBT to  p,p′-dimercaptoazobenzene (DMAB) has been intensively investigated\cite{dong20114}. Both Ag NP-decorated and bare graphene NRs were exposed to the solution of 4-NBT, and the resulting Raman spectra are shown in Figure~5g. Without the NPs only the G-mode of the graphene NRs can be observed. However, edge decorated NPs not only enable surface-enhanced Raman spectroscopy (SERS) signal, but also induce the desired photocatalytic reaction of 4-NBT into DMAB, as evident from the appearance of the DMAB characteristic Raman mode at $\sim$1440~cm$^{-1}$ and $\sim$1390~cm$^{-1}$ .These Raman modes are related to ag$_1$$_6$ and ag$_1$$_7$ vibrations of N=N of DMAB\cite{dong20114}. Such a photocatalytic reaction on nanoribbons decorated with plasmonic nanoparticles shows its future potential towards photocatalytic applications. In further studies, we will focus on employing 2D material-based NRN-FETs combined with edge-specific decorated plasmonic NPs, gaining an additional "knob" via gate biasing. Such coupled mixed-dimensional plasmonic systems can be utilised in gate-controlled photocatalytic reactions, tunable SERS sensors, and high-sensitivity optoelectronic devices, however such experiments would go beyond the scope of this study.

\par$ $\par

\textbf{Conclusion}\par

We propose a method to fabricate nanoribbon networks starting from arbitrary 2D materials, including WS$_2$ and WSe$_2$ NRs, and their heterostructures, which to the best of our knowledge were not demonstrated until now. The method allows achieving NR widths below 20~nm while also enabling a straight forward integration of the 2D material NRNs into high-performance FETs. Further, via the proper choice of the self-aligned molecular masks control of the NR direction with respect to the crystallographic high-symmetry directions is achieved. Examined TMDC nanoribbon network FETs exhibit band transport, maintain high carrier mobility values, clear off-states, high ON-state currents, and importantly stable operation over a large number of sweeping cycles.

Bridging between top-down and bottom-up approaches, our method provides high-quality NR connections (nodes) that do not act as scattering centers (high resistivity points), as proven by \textit{in~operando} KPFM of graphene-NRN FETs. Further, using graphene-NRN FETs we show bi-modal switching of the transfer curves which has been theoretically predicted, and thus far demonstrated only for graphene edges\cite{caridad2018graphene}. By Raman spectroscopy we have observed that MoS$_2$ ribbons with the different growth directions exhibit Raman anisotropy.\cite{wu2016monolayer} 

Lastly, the high edge-to-surface ratios of our NRNs allowed us to selectively decorate the edges with plasmonic nanoparticles. These hybrid mixed-dimensional systems can provide a platform for next generation optoelectronic and plamsonic sensing devices due to the flexibility provided by our method for size tuning of the nanoribbons and the applicability of the process to heterostructures and vertical 2D material p-n junctions.

\par$ $\par

\textbf{\textcolor{black}{Methods}}

$ $\par

\textbf{\textcolor{black}{2D materials, organic masks, and device fabrication:}}
Flakes of 2D materials were mechanically exfoliated from bulk crystal and transferred onto 300~nm SiO$_2$/Si substrate using commercially available Nitto tape and polydimethylsiloxane (Gel-Pak-DLG-X4). Monolayer and few layer flake thicknesses were identified via optical contrast, PL, and Raman measurements. Graphite flakes {10-50~nm} thick kish graphite) were then transferred on 2D materials as electrodes to make device channels. 
6P and DHTA7 nanostructures were grown on devices/flakes by hot wall epitaxy. The growth procedures were adopted from Ref.\cite{matkovic2019light,vasic2018molecules}. MoS2 triangular flakes were grown from solution-based CVD at atmospheric pressure similarly to the procedure in Ref.\cite{kang2019study}. The liquid Mo precursors used were NaMo and AHM in 1:1 ratio and dissolved in ultra-pure water at concentration of 200ppm.\par

\textbf{\textcolor{black}{Reactive Ion Etching:}}
The reactive ion etching process was developed using Oxford Plasma 80 plus RIE system. For all devices the forward power was kept at 80~W with an oxygen flow of 50~sccm under a pressure of 40~mTorr. Etching time was optimized according to the thickness of 2D materials.\par

\textbf{\textcolor{black}{Electrical characterization:}}
Electrical characterization of the flake- and NRN-FETs were done using Keithley 2636A Source-Meter attached to the Instec probe station. The samples were contacted with Au coated Ti electrical cantilever microprobes. Low temperature electrical measurements were performed using liquid nitrogen on a silver plate for thermal uniformity. The temperatures were monitored via mK2000 temperature controller connected to the probe station with a temperature resolution of 0.01K.\par

\textbf{\textcolor{black}{{AFM and FM-}KPFM Measurements}}
AFM and {FM-}KPFM measurements were performed using Horiba/AIST-NT Omegascope AFM system. Aseylec probes were employed (spring constant $\sim$42~N/m, resonant frequency $\sim${70}~kHz, tip radius below 30~nm). For width measurements "Nanosensors" probes were used (spring constant of 10$~$-$~$130N/m, resonant frequency $\sim$ 300 kHz and tip radius of 2 nm). For \textit{in-operando} {FM-}KPFM experiments, the graphene-NRN-FETs were controlled by a Keithley 2636A sharing the same ground with the KPFM-setup. {FM}-KPFM measurements were carried out in a two-pass mode, with the probe lifted by 12~nm in the second pass. Topography and CPD images were processed in the open-source software Gwyddion v2.56. For topography images zero-order line filtering was applied and leveling of the base plane. For CPD images only zero order line filtering was applied.\par

\textbf{{Micro-PL Measurements:}}
All micro-PL and Raman measurements were performed using Horiba LabRam HR Evolution confocal Raman spectrometer using 600 lines/mm and 1800 lines/mm gratings. A 532~nm laser source was used to excite the samples with an excitation power of 0.1 - 3.2 mW. The laser spot was focused by a 100$\times$, 0.9~NA objective.\par

\textbf{{NP edge-decoration and photocatalysis experiments:}}
Ag deposition on graphene nanoribbon was carried out by photodeposition method. 10 µl of 1 mM AgNO3 was dropped on the sample. Thin glass was placed on top of the sample for ease of finding the region of interest. Red laser (633 nm) and objective 100x were used to irradiate the sample. Laser power, laser scanning speed and area were optimized to control the size of Ag NPs. For photocatalytic experiment, 0,1 mM 4-NBT with water to ethanol ratio 50:50 was prepared. Nanoribbon networks decorated with Ag NPs was immersed in this solutions overnight. After that, sample was washed and Raman spectra were recorded with NT-MDT Raman spectroscope.
\par$ $\par

\par
$ $\par
$ $\par

{ \textbf{Acknowledgements:}}{  The authors would like to thank Prof. Roman Gorbachev from the University of Manchester and Prof. Jose$'$ Manuel Caridad from University of Salamanca for their useful input in improving the manuscript. This work is supported by the Austrian Science Fund (FWF) under grants no. I4323-N36 and Y1298-N, and by the Russian foundation for basic research under the project no. 19-52-14006. K.W. and T.T. acknowledge support from the JSPS KAKENHI (Grant Numbers 19H05790, 20H00354 and 21H05233). A.S. and M.K. acknowledge support from the European Regional Development Fund for the “Center of Excellence for Advanced Materials and Sensing Devices” (No. KK.01.1.1.01.0001). Also, the  bilateral Croatian-Austrian project funded by Croatian Ministry of Science and Education and the Centre for International Cooperation and Mobility (ICM) of the Austrian Agency for International Cooperation in Education and Research (OeAD-GmbH) under project HR 02/2020 is acknowledged. Further, the  bilateral French-Austrian project funded bythe Ministère de la Recherche et des Nouvelles Technologies (Amadeus PHC under project no. 42333PL, France) and the Centre for International Cooperation and Mobility (ICM) of the Austrian Agency for International Cooperation in Education and Research (OeAD-GmbH) under project FR 12/2019 is acknowledged.}
\par$ $\par

{ \textbf{Additional information}}\par
{\scriptsize \textbf{Supporting Information:} The online version contains the supplementary information.}\par

{\scriptsize \textbf{Competing financial interests:} The authors declare no competing financial interests.}\par

{\scriptsize \textbf{Keywords:} Nanoribbon, transistors, ferroelectrics, semiconductors,kelvin probe force microscopy, self assembly, nanoparticle}\par

\end{document}